\documentclass[10pt,conference]{IEEEtran}
\usepackage{amsmath,amssymb,mathrsfs}
\usepackage{graphics,epsfig}
\usepackage{graphicx}
\usepackage{psfrag}
\usepackage{subfigure}
\usepackage{euscript}
\newtheorem{note}{NOTE}

\newtheorem{thm}{Theorem}
\newtheorem{lemma}{Lemma}

\newtheorem{defn}{Definition}

\newtheorem{ex}{Example}

\title{Multigroup-Decodable STBCs from Clifford Algebras }

\vspace{1.00cm}

\author{\authorblockN{Sanjay Karmakar and B. Sundar Rajan}
\authorblockA{Department of ECE, Indian Institute Of Science\\
Bangalore, India-560012 \\
Email: \{sanjay,bsrajan\}@ece.iisc.ernet.in}}

\date{}

\begin{document}
\maketitle

\begin{abstract}
A Space-Time Block Code (STBC) in $K$ symbols (variables) is called $g$-group decodable STBC if its maximum-likelihood  decoding metric can be written as a sum of $g$ terms such that each term is a function of a subset of the $K$ variables and each variable appears in only one term. In this paper we provide a general structure of the weight matrices of  multi-group decodable codes using Clifford algebras. Without assuming that the number of variables in each group to be the same, a method of explicitly constructing the weight matrices of  full-diversity, delay-optimal $g$-group decodable codes is presented for arbitrary number of antennas. For the special case of $N_t=2^a$ we construct two subclass of codes: (i) A class of $2a$-group decodable codes with rate $\frac{a}{2^{(a-1)}}$, which is, equivalently, a class of  Single-Symbol Decodable codes, (ii) A class of $(2a-2)$-group decodable with rate $\frac{(a-1)}{2^{(a-2)}}$, i.e., a class of Double-Symbol Decodable codes. Simulation results show that the DSD codes of this paper perform better than previously known Quasi-Orthogonal Designs.
\end{abstract}


\section{Preliminaries and Introduction}
An $N_t\times N_t$ linear dispersion STBC~\cite{HaH} with $K$ real
variables, $x_1, x_2, \cdots,  x_K$ can be written as
\begin{equation}
\label{ldstbc1} 
\mathbf{S}(X)=\sum_{i=1}^{K}x_iA_i
\end{equation}
where $A_i\in \mathbb{C}^{N_t\times N_t}$ and $X=[x_1,x_2, \cdots x_K]\in \mathbb{R}^{1\times K}$ ($\mathbb{C}$ and $\mathbb{R}$ denote respectively the complex and the real field). Now if $
X\in\mathscr{A},$ a finite subset of $\mathbb{R}^K,$ then assuming that perfect channel state information is available at the receiver the maximum likelihood (ML) decision rule minimizes the metric, 
\begin{equation}
\label{mlmetric} 
\mathbf{M(S)} \triangleq \min_{\mathbf{S}}~ tr({(\mathbf{Y-SH})}^{\mathbf{H}}(\mathbf{Y-SH})) = {\parallel
\mathbf{Y}-\mathbf{SH} \parallel}^2.
\end{equation}
It is clear that, in general, ML decoding requires $|\mathscr{A}|$ number of
computations, one for each codeword. Suppose we partition the set of 
weight matrices of the above code into $g$ groups, the $k$-th group 
containing $n_k$ matrices, and  also the information symbol
vector as, $X=[X_1,X_2, \dots X_g]$, where $X_k=[x_{j_k+1},x_{j_k+2}
\cdots x_{j_k+n_k}]$, $j_1=0$  and $j_k=\sum_{i=1}^{k-1}n_i, ~k=2,\cdots,g.$ Now, 
$\mathbf{S}(X)$ can be written as,
\begin{eqnarray*}
\mathbf{S}(X)=\sum_{k=1}^{g} \mathbf{S}_k(X_k),~~~ \mathbf{S}_k(X_k)=\sum_{i=1}^{n_k}x_{j_k+i}A_{j_k+i}.
\end{eqnarray*}
If the weight matrices of \eqref{ldstbc1} are such that,
\begin{equation}
\label{ldstbc_gp}
\mathbf{S}^H(X)\mathbf{S}(X)=\sum_{k=1}^{g}\mathbf{S}^H_k(X_k)\mathbf{S}_k(X_k)
\end{equation}
and $X_k\in \mathscr{A}_k\subset \mathbb{R}^{n_k}, \forall k$ take values independently then using \eqref{ldstbc_gp} in \eqref{mlmetric} we get,
\begin{equation}
\label{mlmetric1} 
\mathbf{M(S)} =\sum_{k=1}^{g} {\parallel
\mathbf{Y}-\mathbf{S}_k(X_k)\mathbf{H}
\parallel}^2-(g-1){\parallel Y \parallel}^2
\end{equation}
from which it follows that minimizing the metric in \eqref{mlmetric} is equivalent to
minimizing,
\begin{equation*}
\label{mlmetric2} \mathbf{{M(S)}_k} = {\parallel
\mathbf{Y}-\mathbf{S}_k(X_k)\mathbf{H}
\parallel}^2
\end{equation*}
for each $1 \leq k\leq g$ individually.
\begin{defn}
\label{defn_ggp} A linear dispersion STBC as in \eqref{ldstbc1} is
called $g$-group decodable if its decoding metric in
\eqref{mlmetric} can be simplified as in \eqref{mlmetric1} and the
information symbols in each group takes values independent of 
information symbols in other groups.
\end{defn}
It is easily that the ML decoding of a $g$-group decodable code requires
only $\sum_{k=1}^{g}|{\mathscr{A}}_k|$ computations which in general
is much smaller than $|\mathscr{A}|=\prod_{k=1}^{g}{\mathscr{A}}_k.$ Single-Symbol Decodable (SSD) and Double-Symbol Decodable (DSD) codes have been studied extensively \cite{KaR1},\cite{YGT2},\cite{TiH}. Note that a SSD code in $K$ variables is nothing but a $K$-group decodable code and a DSD code is nothing but a $\frac{k}{2}$-group decodable code. For example, a $4\times 4$ SSD code \cite{WWX1},\cite{YGT2},\cite{KhR1} or CIOD \cite{KhR1} is
4-group decodable code, each group containing two real information
symbols to be decoded together. 

 In this paper we study general $g$-group decodable codes using Clifford algebras. The main contributions of this paper can be summarized as follows: 
\begin{itemize} 
\item All known results for $g$-group decodable codes so far, including the most recent one \cite{YGT1} study $g$-group decodable codes in which each group contain the same number of information symbols. In this paper we give a general algebraic structure of the weight matrices of $g$-group decodable codes, where different groups can have different number of information symbols to be decoded together.
\item
Recently $g$-group decodable codes, for $g=4,6,8$ have been reported
in \cite{YGT1}. Due to the recursiveness of the reported construction procedure  delay optimal codes for number of transmit antennas, which
is not a power of 2 are not obtainable from the techniques of \cite{YGT1}. 
Whereas due to our general construction procedure being non-recursive, we can construct delay-optimal $g$-group
decodable codes even for $N_t\neq 2^a$. Example \ref{example2} of
Section \ref{sec3} is one such code. 

\item An analytic expression for the diversity product of our code is given (Section \ref{sec4}), using which the full diversity property of the codes is established.
\end{itemize}

The remaining part of the paper is organized as follows: In Section \ref{sec2} we present construction of weight matrices of a class of linear dispersion codes which will facilitate the design of multi-group decodable codes in the following section. In Section \ref{sec3} we present our explicit construction of $g$-group decodable codes for all values of $g$. A closed form expression for the diversity product of our codes is obtained in Section \ref{sec4} and in Section \ref{sec5} we present SSD and DSD codes obtainable from our construction of Section \ref{sec3} along with simulation results for one such code.  

The proofs of all the theorems and lemmas have been omitted due to lack of space.
\section{General Structure of Multigroup Decodable codes}
\label{sec2}
In this section, we describe a construction of weight matrices of a linear dispersion code which will greatly facilitate the design of multi-group decodable codes subsequently. Let there be $(g+1)$ number of collection of matrices $G_0,G_1,\cdots G_g$, where
\begin{eqnarray}
\label{gr_matrices} 
\begin{array}{ccc} 
G_0&=&\{A_{0,k}\in \mathbb{C}^{m\times m}\,\lvert \,k=1,2,\cdots g\}\\
G_1&=&\{A_{1,i_1}\in \mathbb{C}^{n\times n} \,\lvert \, i_1=1,2,\cdots n_1\}\\
\vdots & & \vdots \\
G_g&=&\{A_{g,i_g}\in \mathbb{C}^{n\times n}\,\lvert \, i_g=1,2,\cdots n_g\}
\end{array}
\end{eqnarray}
Now from the above set of matrices we form a new set of  following matrices, which can be used as weight matrices of LD codes subsequently:
\begin{equation}
\label{wt_matrices} 
W=\{A_{0,k}\otimes A_{k,i_k}\,\lvert\,1\leq
k\leq g; 1\leq i_k\leq n_k\}
\end{equation}

Let  $K=|W|=\sum_{k=1}^{g}n_k$ be the cardinality of $W$ and the information bits to be transmitted is mapped to a real vector $X=[x_1, \cdots x_K]\in \mathscr{A}\subset \mathbb{R}^{K}$, where $\mathscr{A}$ is finite. Then we construct the corresponding STBC as follows,
\begin{eqnarray}
\label{ldstbc2} 
\mathbf{S}(X)=\sum_{k=1}^{g}\sum_{i=1}^{n_k}x_{j_k+i}A_{0,k}\otimes A_{k,i}=\sum_{k=1}^{g}\mathbf{S}_k(X_k); \\ \mathbf{S}_k(X_k)=\sum_{i=1}^{n_k}x_{j_k+i}A_{0,k}\otimes A_{k,i}
\end{eqnarray}
where $X=[X_1,X_2, \cdots X_g]$ and $X_k=[x_{j_k+1},\cdots x_{j_k+n_k}]\in \mathbb{R}^{n_k},$  $ 1\leq k \leq g$.
\begin{thm}
\label{theorem1} 
The linear dispersion code given in \eqref{ldstbc2}
is a $g$-group decodable code, the $k$th group involving $n_k$
information symbols  of $X_k$, if the following conditions are
satisfied,
\begin{eqnarray*}
\label{thm1I}
X_1,X_2,\cdots X_g  \textrm{    are mutually independent} \\
\label{thm1II} 
A_{0,i}^HA_{0,j}+A_{0,j}^HA_{0,i}=0,\forall
1\leq i\neq j\leq g\\
\label{thm1III} 
A^H B=B^H A,  ~~~\forall A\in G_i,B\in G_j,1\leq
i\neq j \leq g.
\end{eqnarray*}
\end{thm}

\begin{thm}
\label{theorem2} Suppose the $(g+1)$ set of matrices of
\eqref{gr_matrices} satisfy Theorem \ref{theorem1} and moreover, the
weight matrices corresponding to  $G_k, 1\leq k\leq g$ can be subdivided into $g_k$
subgroups, i.e.,$G_k=G_{k,1}\cup G_{k,2} \cup, \cdots, \cup G_{k,g_k}$, such
that,
\begin{eqnarray*}
\label{thm2I} A_k^H B_k+B_k^H A_k=0, ~~ \forall 1\leq k \leq g, \\ 
\mbox{  where   } A_k\in G_{k,i}, ~~ B_k\in G_{k,j}, ~~~ 1\leq i\neq j \leq g_k
\end{eqnarray*}
and the corresponding information vectors, $X_{k,1}\cdots X_{k,g_k}
$ are independent, where $X_k=[X_{k,1}, X_{k,2}\cdots X_{k,g_k}]$
and $X_{k,l}\in \mathbb{R}^{|G_{k,l}|}$.  Then the ML decoding of
$X_k$ can further be separated into $g_k$ subgroups, for each $1\leq
k \leq g$.
\end{thm}

If the collection of matrices in \eqref{gr_matrices} of the STBC given in \eqref{ldstbc2} satisfies Theorem \ref{theorem1} and Theorem \ref{theorem2} simultaneously, then the code is $(\sum_{k=1}^{g}g_k)$-group decodable.

\section{Explicit construction of Multigroup Decodable Codes}
\label{sec3} 
In this section we construct $g$-group decodable codes for any value of  $g.$
\begin{thm}
\label{corollary1} Let $\widetilde{G}$ be a set of $n\times n$
mutually commuting Hermitian complex matrices and $G_0$ is a set of weight
matrices such that, for any $A,B\in G_0, A^HB+B^HA=0$. Now if we
choose $G_1=G_2=\cdots G_g=\widetilde{G}$, where $g=|G_0|$ and
construct the weight matrices as in \eqref{wt_matrices} and further
construct a STBC as in \eqref{ldstbc2}, then the resulting code
will be a $g$-group decodable STBC with rate,
\begin{equation}
\label{real_rate} 
R_r=\frac{|\widetilde{G}||G_0|}{mn}
\end{equation}
real information symbols per channel use.
\end{thm}

Now if we want to construct a $g$-group decodable code, according to
Theorem \ref{corollary1} above we need to select the collection of
matrices $G_0$ with cardinality at least $g$. But from
\eqref{real_rate} the rate is dependent on choice of $G_0$ through
$g=|G_0|$ and $m$. So for larger rate it is better to choose $m$ as
small as possible as $g$ is fixed. This is a very hard problem in
general to solve. So  we will assume $G_0$ to be a collection of unitary matrices since the answer to the above question is available in
\cite{TiH} for these cases. The answer is,  for $g$ matrices the minimum value of $m$ is given by  $m=2^{\lfloor \frac{g-1}{2}\rfloor}$. 
Note that with this result, we have for every $g$,  a $g$-group decodable
code for $N_t=2^{\lfloor \frac{g-1}{2}\rfloor}$ transmit antennas in
\cite{TiH}. Here $\widetilde{G}=\{1\}$ is the trivial set. Now suppose we want a $g$-group decodable code for $N_t$ transmit antennas,
where $N_t=m(=2^{\lfloor \frac{g-1}{2}\rfloor})n, ~~ n\geq 2$. Then $\widetilde{G}$ must contain $n\times n$ Hermitian, mutually
commuting complex matrices, according to Theorem \ref{corollary1}.
But again from \eqref{real_rate} the rate of the code (that we are
going to construct) depends on the choice of $\widetilde{G}$ through
$|\widetilde{G}|$ and $n$. As $n$ is fixed ($N_t=mn$ is given and we
have found $m$ during the choice of $G_0$), we need to make the cardinality
of $\widetilde{G}$ as large as possible. Again at this stage we will
assume unitarity of the matrices in $\widetilde{G}$. With this
assumption we obtain the following lemma on the cardinality of
$\widetilde{G}$,
\begin{lemma}
\label{comm_set} 
The cardinality of the set $\widetilde{G}$ of
Theorem \ref{corollary1} is $n$ under the unitarity assumption, and
the assumption that the resulting code is uniquely decodable.
\end{lemma}

With this result we see that the code constructed following
Theorem \ref{corollary1} will be of rate $R_r=\frac{|\widetilde{G}||G_0|}{mn}=\frac{g}{m}$ real information symbols per channel use.
Note that the construction suggested in the description above is far from general and the weight matrices of the codes constructed by this method will be unitary.\\

To explain the construction of the set $G_0$ we need irreducible
matrix representation of Clifford Algebra.
\begin{defn}
The Clifford algebra, denoted by $CA_L$, is the algebra over the
real field $\mathbb R$ generated by $L$ objects $\gamma_k,
~~~k=1,2,\cdots,L$ which are anti-commuting, ($\gamma_k\gamma_j =
-\gamma_j \gamma_k, ~ \forall k\neq j,$) and squaring to $-1$,
($\gamma_k^2 = -1 ~~ \forall k=1,2,\cdots,L$).
\end{defn}

Let
\begin{equation*}
\label{paulimatrices} \sigma_1 =\left[ \begin{array}{rr}
0 & 1 \\
-1 & 0
\end{array}
\right], \sigma_2 =\left[ \begin{array}{rr}
0 & j \\
j & 0
\end{array}
\right] \mbox{  and  } \sigma_3 =\left[ \begin{array}{rr}
1 & 0 \\
0 & -1
\end{array}
\right],
\end{equation*}
$~~~ \sigma_4=-j\sigma_2$ and $~~$
$A^{\otimes^{m}} = \underbrace{A\otimes A\otimes A \cdots \otimes A }_{m~~times  } $. \\
From \cite{TiH} we know that the representation $R(\gamma_j), ~j=1,2,\cdots,L$ of the generators of
$CA_{2a+1}$ are obtainable in terms of $\sigma_i, ~~ i=1,2,3,4.$ and explicitly shown in \cite{TiH}.
%
\subsection{Construction of $G_0$}
\label{g0_construction} If $g$ is even, say $(g-1)=(2a+1)$, find the
irreducible representation of ${CA}_{g-1}$ as described in \cite{TiH}. Then our required set $G_0$ is,
\begin{equation*}
\label{g0matrices} G_0=\{R(\gamma_0)=I_{m\times
m},R(\gamma_1),R(\gamma_2),\cdots R(\gamma_(g-1))\}
\end{equation*}
Here $R(\gamma_i)\in \mathbb{C}^{m\times m}$ and
$m=2^{\lfloor{\frac{g-1}{2}}\rfloor}$. Similarly for say, $g=2a+1$
odd we find the irreducible representation of ${CA}_g$ and add to this set the identity matrix. Thus we will get $g+1$ matrices. We can use any
$g$ of them (or we can use all $(g+1)$ of them and consider any two
groups as a single one, this way we can increase the rate).

\subsection{construction of $\widetilde{G}$}
Lemma \ref{comm_set} above suggest a construction method of the set
$\widetilde{G}$. Following that method, for a given $n$ we
will first find  $n$ linearly independent vectors $b_i\in
{\{+1,-1\}}^n, 1\leq i\leq n$. Then we will choose an $n\times n$
unitary matrix. The choice of this matrix is important as explained in Note \ref{choice_of_U} below. Now we can construct the set as follows,
\begin{equation*}
\label{gimatrices} \widetilde{G}=\{U Diag(b_i)U^H | i=1,2, \cdots
n\}.
\end{equation*}

\begin{note}
\label{choice_of_U} Note that in the above construction
$U=I_{n\times n}$ may be a choice. But then the resulting matrices
will be diagonal and will contain a large number of zero entries. This will
lead to a large PAPR of the code. So $U$ need to be chosen in such a
way that the matrices in $\widetilde{G}$ have as small number of
zero entries as possible.
\end{note}
As an example we will construct below a $4$-group decodable code for
$N_t=6$ transmit antennas which is delay optimal. As mentioned
earlier this code can't be obtained following the approach of
\cite{YGT1}. This code also has rate 1.
\begin{ex}
\label{example2} According to the construction procedure of ${G}_0$
described above we choose,
\begin{equation*}
\label{ex2_g0matrices} 
G_0=\{I_{2},\sigma_1,\sigma_2,j\sigma_3\}.
\end{equation*}
For this example we don't take the trouble to find an appropriate $U$
as explained in Note \ref{choice_of_U}. Instead we choose
$U=I_{3\times 3}$. Thus our set $\widetilde{G}$ is,
\begin{eqnarray*}
\label{ex2_gimatrices}
G_i = \widetilde{G} = & \{Diag([1,1,1]),Diag([1,1,-1]),\\
& Diag([-1,1,1])\}, i=1,2\cdots 4.
\end{eqnarray*}
With this set of matrices and an information vector $X=[x_1, \cdots
x_{12}]$, we construct the STBC according to \eqref{ldstbc2} as given in \eqref{66code} at the top of the next page,
\begin{figure*}
{\small
\begin{equation}
\label{66code} \mathbf{S}(X)=\left[ \begin{array}{cccccc}
(z_1+z_2-z_3) &
0 & 0 & (z_4+z_5-z_6) & 0 & 0 \\
0 & (z_1+z_2+z_3) & 0 & 0 & (z_4+z_5+z_6) & 0 \\
0 & 0 & (z_1-z_2+z_3) & 0 & 0 & (z_4-z_5+z_6) \\
-(z_4+z_5+z_6)^{\ast} & 0 & 0 & (z_1+z_2-z_3)^{\ast} & 0 & 0 \\
0 & -(z_4+z_5+z_6)^{\ast} & 0 & 0 & (z_1+z_2+z_3)^{\ast} & 0 \\
0 & 0 & -(z_4+z_5+z_6)^{\ast} & 0 & 0 & (z_1-z_2+z_3)^{\ast} \\
\end{array} \right]
\end{equation} \\ \hrule
}
\end{figure*}
where $z_1=x_1+jx_{10}, z_2=x_2+jx_{11},z_3=x_3+jx_{12},
z_4=x_4+jk_7,z_5=x_5+jk_8,z_6=x_6+jk_9$. In the next section we  prove  that 
 this code is of full-diversity by showing that every code constructed according to the Theorem \ref{corollary1} achieve full diversity.
\end{ex}

\section{Diversity Product of Multi-group decodable codes}
\label{sec4}

Let $\mathbf{S}(X)$ be a $g$-group decodable code, constructed
according to Theorem \ref{corollary1} where $X=[X_1,\cdots X_g],
X_k\in\mathscr{A}_k\subset\mathbb{R}^{n},\forall 1\leq k\leq g$.
Let's also denote $\mathscr{A}_1\times \cdots \times
\mathscr{A}_g=\mathscr{A}$. Now suppose, $X\neq\widetilde{X}\in
\mathscr{A}$ and $\Delta X =X-\widetilde{X}$. Then,
\begin{equation*}
\label{dp1}
\begin{array}{l}
\mathbf{S}(X)-\mathbf{S}(\widetilde{X})=\mathbf{S}(\Delta X) \\
=\sum_{k=1}^{g} \sum_{i=1}^{n}\Delta x_{(k-1)n+i}A_{0,k}\otimes A_i, \\
~~~~~~~~~A_{0,k}\in G_0, A_i\in \widetilde{G}
\end{array}
\end{equation*}
and
\begin{equation}\label{dp2}
\begin{array}{l}
\mathbf{S}^H(\Delta X)\mathbf{S}(\Delta X)\\
=\sum_{k=1}^{g}\Bigg\{{(\sum_{i=1}^{n}\Delta x_{(k-1)n+i}I_{m\times
m}\otimes A_i)}^H \\
  ~~~~~~~~~~~~~~~~~~~    (\sum_{i=1}^{n} \Delta x_{(k-1)n+i}I_{m\times
m}\otimes A_i)\Bigg\}.
\end{array}
\end{equation}
Now according to construction,
\begin{equation*}
\label{struc_of_g}
\widetilde{G}=\{A_i=U Diag(b_i)U^H,
i=1,2 \cdots  n\}
\end{equation*}
using which in \eqref{dp2} we get,
\begin{equation}
\label{dp3}
\begin{array}{l}
\mathbf{S}^H(\Delta X)\mathbf{S}(\Delta X)\\
=I_{m\times m}\otimes U \sum_{k=1}^{g}  \Big\{ \sum_{i=1}^{n} \Delta
x_{(k-1)n+i} I_{m\times m} \\
~~~~~~~~~~~~~~ \otimes Diag(b_i){\Big\}}^2 I_{m\times m}\otimes
U^H.
\end{array}
\end{equation}
Now let,
\begin{eqnarray*}
\label{tr1}
\begin{array}{l}
Y_k=[y_{(k-1)n+1},y_{(k-1)n+2}, \cdots
y_{(k-1)n+n}] \\
=\underbrace{\frac{1}{c}[b_1^T,b_2^T,\cdots
b_n^T]}_{T}{[x_{(k-1)n+1},x_{(k-1)n+2}, \cdots x_{(k-1)n+n}]}^T.
\end{array}
\end{eqnarray*}
Then, if $X_k\in \mathscr{A}_k$ then $Y_k\in \mathscr{B}_k\subset\mathbb{R}^n$. Here $c$ is chosen so that the average energy of both the constellations $\mathscr{A}_k$ and $\mathscr{B}_k$ is same. And as the transform $T$ is non singular, there is a one-to-one correspondence between the points in $\mathscr{A}_k$ and $\mathscr{B}_k, \forall k$. Now using this in \eqref{dp3} we get,
\begin{equation*}\label{dp4}
\begin{array}{ll}
\mathbf{S}^H(\Delta X)\mathbf{S}(\Delta X) & \\
=I_{m\times m}\otimes U \sum_{k=1}^{g}c^2 \Big\{ I_{m\times m} & \\
 ~~~~~~~~~~~~~~~~~\otimes Diag(\Delta Y_k){\Big\}}^2 I_{m\times m}\otimes U^H & \\
=c^2 I_{m\times m} \otimes U  I_{m\times m} \otimes
Diag\Bigg(\Big[\sum_{k=1}^{g}{\Delta y_{(k-1)n+1}}^2,&\\
\sum_{k=1}^{g}{\Delta y_{(k-1)n+2}}^2, \cdots
\sum_{k=1}^{g}{\Delta y_{(k-1)n+n}}^2\Big]\Bigg) \Big\} I_{m\times m}\otimes U^H
\end{array}
\end{equation*}
Here $Y=[Y_1,Y_2,\cdots Y_g]$ and $\Delta Y_k, \forall k$ is defined similarly as $\Delta X_k$. Then 
\begin{equation*}\label{dp5}
\det \left(\mathbf{S}^H(\Delta X)\mathbf{S}(\Delta
X)\right)=c^{2N_t}{\left[\prod_{i=1}^{n}\left(\sum_{k=1}^{g}
\Delta y_{(k-1)n+i}^{2}\right)\right]}^m.
\end{equation*}
Let's now define,
\begin{equation*}
\label{dp6}
\begin{array}{l}
DP1\triangleq \min_{\Delta X \neq 0} \det \left(\mathbf{S}^H(\Delta
X)\mathbf{S}(\Delta X)\right) \\
=\min_{\Delta Y \neq 0}
c^{2N_t}{\left[\prod_{i=1}^{n}\left(\sum_{k=1}^{g}
y_{(k-1)n+i}^{2}\right)\right]}^m \\
=\min_{\Delta Y_k \neq 0,\textrm{for all}~ k}
c^{2N_t}{\left[\prod_{i=1}^{n}\underbrace{\left(\sum_{k=1}^{g}
y_{(k-1)n+i}^{2}\right)}_{p_i}\right]}^m.
\end{array}
\end{equation*}
Notice that for all $i, p_i$ is a sum of positive numbers and hence the above expression is minimized when $\Delta Y_k\neq 0$ for only one value of $k$. Hence,
\begin{equation*}
\label{dp7b}
DP1=\min_{1\leq k\leq g | \Delta Y_k \neq 0}
c^{2N_t}{\left[\prod_{i=1}^{n}\left(
y_{(k-1)n+i}^{2}\right)\right]}^m.
\end{equation*}

The last expression is same for all $k$. So we assume that all $Y_k$
takes values from the same $n$-real dimensional constellation, i.e.,
$Y_k\in \mathscr{A}_y \forall k$. This actually means that we are assuming that all $X_k$ takes their values from the same $n$-real dimensional constellation, i.e., $\{\mathscr{A}_1= \mathscr{A}_2\cdots =
\mathscr{A}_g=\mathscr{A}_x\}$(say). So without loss of generality we
assume $k=1$ and have,
\begin{equation*}
\label{dp8}
 DP1=\min_{\Delta Y_1 \neq 0}
c^{2N_t}{\left(\prod_{i=1}^{n}\Delta y_i\right)}^{2m}.
\end{equation*}
Hence the diversity product of the code in \cite{SuX} is,
\begin{eqnarray}
\label{dp}
DP\triangleq \min_{\Delta X \neq 0} \frac{1}{2\sqrt{N_t}}{\det
\left(\mathbf{S}^H(\Delta X)\mathbf{S}(\Delta X)\right)}^{2N_t} \nonumber \\
\label{dp0}
=\min_{\Delta Y_1 \neq 0}\frac{c}{2\sqrt{N_t}}
{\left(\prod_{i=1}^{n}\Delta y_i\right)}^{\frac{1}{n}}.
\end{eqnarray}
Hence from \eqref{dp0} we conclude that the diversity
product of the codes constructed following Theorem \ref{corollary1}
is a function of the CPD of the finite subset $\mathscr{A}_y$ of the $n$-real dimensional vector space, $\mathbb{R}^n$, which is actually the linearly transformed version of $\mathscr{A}_x$ from which all $X_k$ takes their values. Now our strategy will be to select a $\mathscr{A}_y$ with its CPD being maximal\cite{Vit}. Then we apply a linear transform $T^{-1}$ to get $\mathscr{A}_x$. Now allow $X_k\in \mathscr{A}_x ,\forall k$. Thus the resulting code will achieve the maximal (non-zero) diversity product.
\section{Construction of SSD and DSD codes}
\label{sec5}
In this section we construct SSD codes \cite{KhR1} and DSD codes using a modified version of the construction  described in Section \ref{sec3}. 
Towards this end, we first give an
alternative construction of the set $\widetilde{G}$ for $N_t=2^a,
a\in \mathbb{N}$.

\subsection{Alternative Construction of $\widetilde{G}$}
\label{alternate_construction} 
For $N_t=2^a, a\in \mathbb{N}$ we take the matrices $\{R(\gamma_1),R(\gamma_2),\cdots
R(\gamma_{(2a+1)})\}$ as given in \cite{TiH}. From this set we construct
$\{\widetilde{A}_1=jR(\gamma_1)R(\gamma_{a+1}),
\widetilde{A}_2=jR(\gamma_2)R(\gamma_{a+2}),\cdots
\widetilde{A}_a=jR(\gamma_a)R(\gamma_{2a})\}$. It can be easily verified that these matrices are commuting and Hermitian. Now from
this we construct the set we require containing $2^a$ matrices as
follows,
\begin{equation} \label{comm_set1}
\begin{array}{l}
\widetilde{G}=\{I_{n\times n}\}\cup\{\pm \widetilde{A}_k|k=1,\cdots
a\}\\
~~~~~~~~~~~~~~~~~~\cup_{j=2}^{a} \{\pm \prod_{i=1}^{j}
\widetilde{A}_{k_i} | 1\leq k_i<k_{(i+1)}\leq a\}
\end{array}
\end{equation}
Note that the matrices in \eqref{comm_set1} are
all distinct, unitary, Hermitian and mutually commuting $n\times n$
complex matrices.

\begin{figure*}
{\footnotesize
\begin{equation}
\label{dsd88_code}
\left[
  \begin{array}{cccccccc}
(x_1+jx_{13}) &(-x_{14}+jx_2)&(x_4+jx_{15})  &(-x_{15}+jx_3) &(x_5+jx_9)    &(-x_{10}+jx_6) &(x_8+jx_{12})  &(-x_{11}+jx_7)  \\
   (x_{14}-jx_2) & (x_1+jx_{13}) &(-x_{15}+jx_3) &(-x_4-jx_{15}) &(x_{10}-jx_6) &(x_5+jx_9)     &(-x_{11}+jx_7) &(-x_8-jx_{12})  \\
   (x_4+jx_{15}) & (x_{15}-jx_3)&(x_1+jx_{13})   &(x_{14}-jx_2)  &(x_8+jx_{12}) &(x_{11}-jx_7)  &(x_5+jx_9)     &(x_{10}-jx_6)  \\
   (x_{15}-jx_3) &(-x_4-jx_{15})&(-x_{14}+jx_2) &(x_1+jx_{13})   &(x_{11}-jx_7) &(-x_8-jx_{12}) &(-x_{10}+jx_6) &(x_5+jx_9)  \\
   (-x_5+jx_9)   &(-x_{10}-jx_6)&(-x_8+jx_{12}) &(-x_{11}-jx_7) &(x_1-jx_{13})  &(x_{14}+jx_2)  &(x_4-jx_{15})  &(x_{15}+jx_3)  \\
   (x_{10}+jx_6) &(-x_5+jx_9)   &(-x_{11}-jx_7) &(x_8-jx_{12})  &-(x_{14}+jx_2)&(x_1-jx_{13})   &(x_{15}+jx_3)  &(-x_4+jx_{15})  \\
   (-x_8+jx_{12})&(x_{11}+jx_7) &(-x_5+jx_9)    &(x_{10}+jx_6)  &(x_4-jx_{15}) &-(x_{15}+jx_3) &(x_1-jx_{13})   &(-x_{14}+jx_2)  \\
    (x_{11}+jx_7)&(x_8-jx_{12}) &-(x_{10}+jx_6) &(-x_5+jx_9)    &-(x_{15}+jx_3)&(-x_4+jx_{15}) &(x_{14}+jx_2)  &(x_1-jx_{13})  \\
  \end{array}
\right]
\end{equation}}\\ \hrule
\end{figure*}


\subsection{SSD codes}
\label{SSD_codes} Suppose we want to construct SSD code for $N_t=2^a,
a\geq 2$ transmit antennas. In other words the codes to be
constructed are $g$-group decodable for some $g$, where each group
contain only two real symbols. This imply that $n=2$. From the above
construction we find, $\widetilde{G}=\{I_{2\times 2}, \sigma_4\}.$ Now from $N_t=mn=2m$ we get the value of $m$. Next we need to find
the set $G_0$. Following the discussion in
Subsection \ref{g0_construction} we can construct the set $G_0$, for
this value of $m$ which is illustrated in the following example.
\begin{ex}
\label{ssd_44_ex} We take, $N_t=4$. Then
$\widetilde{G}=G_i,i=1,2,3,4$ is as described above. For $m=2$ we
get, $G_0=\{I_2,\sigma_1,\sigma_2,j\sigma_3\}$. Next we construct
the STBC according to \eqref{ldstbc2} as,
\begin{equation*}\label{ssd44code}
\left(
  \begin{array}{cccc}
    x_1+jx_7 &  x_2+jx_8 & x_3+jx_5 & x_4+jx_6 \\
    x_2+jx_8 &  x_1+jx_7 & x_4+jx_6 & x_3+jx_5 \\
   -x_3+jx_5 & -x_4+jx_6 & x_1-jx_7 & x_2-jx_8 \\
   -x_4+jx_6 & -x_3+jx_5 & x_2-jx_8 & x_1-jx_7 \\
  \end{array}
\right)
\end{equation*}
which is $4$-group decodable.
\end{ex}
\begin{note}
\label{} In general for any given $N_t=2^a$ number of transmit
antennas, we get a $2a$-group decodable code, with rate
$\frac{a}{2^{(a-1)}}$ complex symbols per channel use. Interestingly
this was reported in \cite{KaR1} as the maximum rate of Unitary Weight SSD
codes.
\end{note}

\subsection{DSD Codes}
\label{DSD_codes} DSD codes can also be viewed as  $g$-group
decodable codes for some $g$, where each group contains two complex
symbol or 4 real symbols to be decoded together, which means that for this
class of codes $n=|\widetilde{G}|=4$. From the construction above in
Subsection \ref{alternate_construction} we get,
\begin{equation}\label{dsd1}
\widetilde{G}=\{I_{2\times 2}\otimes I_{2\times 2}, \sigma_3\otimes
 j\sigma_1,\sigma_1\otimes
 \sigma_2,\sigma_4\otimes
 \sigma_3\}.
\end{equation}
For any given $N_t=2^a$, we find $m=\frac{N_t}{n=4}$. Then following
the construction procedure in Subsection \ref{g0_construction} we
find the set $G_0$, and then construct the STBC according to
\eqref{ldstbc2}.

\begin{ex}
\label{dsd88_ex} Let us take $N_t=8$. Then
$\widetilde{G}=G_i,i=1,2,3,4$ is given by \eqref{dsd1}. For $m=2$ we
get, $G_0=\{I_2,\sigma_1,\sigma_2,j\sigma_3\}$. Next we construct
the LDSTBC according to \eqref{ldstbc2} and is given in
\eqref{dsd88_code} at the top of this page. According to the
construction this is a $4$-group decodable code.
\end{ex}
In general for any given $N_t=2^a$ number of transmit
antennas, we get a $(2a-2)$-group decodable code, with rate
$\frac{(a-1)}{2^{(a-2)}}$ complex symbols per channel use.

\subsection{Simulation Results of DSD codes}
\begin{figure}
\centering
\includegraphics[width=8.0cm,height=8.0cm]{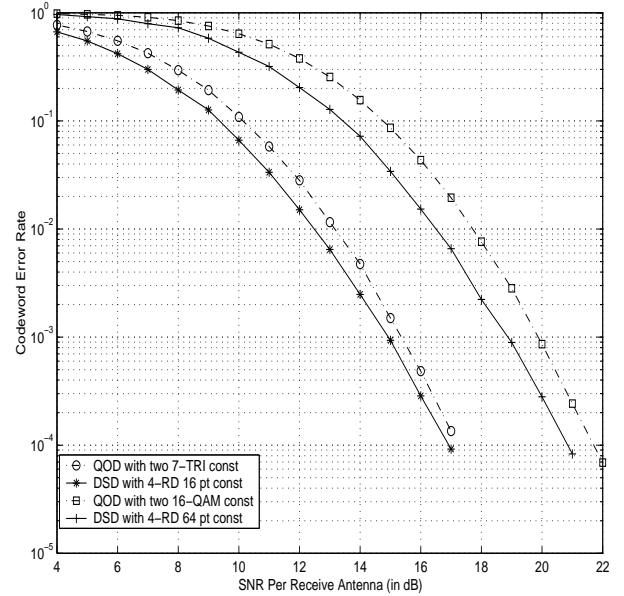}
\caption{Comparison of DSD with the QOD code at bit rate of 2 bits per channel use. }
\label{fig1_eps}
\end{figure}

In Figure \ref{fig1_eps} we have compared the performance of QOSTBC \cite{SuX} and DSD code for $8$-transmit antennas. For QOSTBC we used two 7-ary constellation optimally rotated as in \cite{SuX}. For DSD as $\mathscr{A}_y$, we used a 16-point $4$-real dimensional CPD-optimized constellation. And then obtained $\mathscr{A}_x$ by transforming $\mathscr{A}_y$ by $T^{-1}$. Then we allowed $X_k\in \mathscr{A}_x, \forall k$.

\section*{Acknowledgment}
This work was partly supported by
the DRDO-IISc Program on Advanced Research in Mathematical
Engineering, partly by the Council of Scientific \&
Industrial Research (CSIR), India, through Research Grant (22(0365)/04/EMR-II) and also by Beceem Communications Pvt. Ltd., Bangalore to B.S.~Rajan.

\end{document}